\documentclass[a4paper]{article}
\usepackage{epsfig,amsmath,amsfonts,amssymb,setspace,multirow,textcomp}
\usepackage[T1]{fontenc}
\makeatletter
\renewcommand{\@oddhead}{\textit{} \hfil}
\renewcommand{\@evenfoot}{\hfil \thepage \hfil}
\renewcommand{\@oddfoot}{\hfil \thepage \hfil}
\makeatother

\renewenvironment{thebibliography}[1]{\begin{oldthebibliography}{#1}\setlength{\parskip}{0ex}\setlength{\itemsep}{0ex}}{\end{oldthebibliography}}
\usepackage{enumerate}
\usepackage{caption}
\usepackage{cases}
\usepackage{multicol}
\usepackage{cite}
\usepackage{hyperref}
\usepackage{ulem}
\hypersetup{colorlinks = true, citecolor = {blue}, urlcolor = {blue}, linkcolor={blue}}
\renewcommand{\arraystretch}{1.2}

\begin{document}
\fontsize{11}{11}\selectfont 
\title{
Milky Way Globular Clusters: close encounter rates with each other and with the Central Supermassive Black Hole
}
\author{\textsl{M.\,V.~Ishchenko$^{1}$\footnote{{\href{mailto:marina@mao.kiev.ua}{\tt marina@mao.kiev.ua}}},
M.\,O.~Sobolenko$^{1}$, 
M.\,T.~Kalambay$^{2,3,4}$,
B.\,T.~Shukirgaliyev$^{4,3}$,
P.\,P.~Berczik$^{5,1}$}}
\date{\vspace*{-6ex}}
\maketitle
\begin{center}{\small 
$^{1}$Main Astronomical Observatory, National Academy of Sciences of Ukraine,\\
27 Akademika Zabolotnoho St., 03143, Kyiv, Ukraine\\
$^{2}$Al-Farabi Kazakh National University, 71 Al-Farabi Av., 050040 Almaty, Kazakhstan\\
20A Datun Rd., Chaoyang District, 100101 Beijing, China\\
$^{3}$Fesenkov Astrophysical Institute, 23 Observatory St., 050020 Almaty, Kazakhstan\\
$^{4}$Energetic Cosmos Laboratory, Nazarbayev University, 52 Kabanbay Batyr Av., 010000 Nur-Sultan, Kazakhstan\\
$^{5}$National Astronomical Observatories and Key Laboratory of Computational Astrophysics, Chinese Academy of Sciences,\\
20A Datun Rd., Chaoyang District, Beijing 100101, China}
\end{center}

\begin{abstract}
Using the data from \textit{Gaia} (ESA) Data Release~2 we performed the orbital calculations of globular clusters~(GCs) of the Milky Way. To explore possible collisions between the GCs, using our developed high-order $\varphi$-GRAPE code, we integrated (backwards and forward) the orbits of 119~objects with reliable positions and proper motions. In calculations, we adopted a realistic axisymmetric Galactic potential (\textit{bulge + disk + halo}). Using different impact conditions, we found three pairs of the GCs that likely experienced collisions:  Terzan~3 -- NGC~6553, Terzan~3 -- NGC~6218, Liller~1 -- NGC~6522, Djorg~2 -- NGC~6552 and NGC~6355 -- NGC~6637. 

We analyzed the GCs interaction rates with the central supermassive black hole. Assuming the maximum 100~pc distance criteria for separation between them we estimated 11 close encounter events. From our numerical simulations we estimate the close interaction rate as: at least one event per Gyr with the impact parameter less than 30~pc; and one event per Myr with the impact parameter less than 60~pc. Our calculations show one very close encounter of NGC~6121 with the central SMBH near 5.5~pc (practically direct collision). Based on the extended literature search for the possible progenitor of our selected 11~GCs, we found that most of them have a Milky Way main bulge origin.
\\[1ex]
\end{abstract}

\begin{multicols}{2}
\section*{\sc introduction}
\indent \indent 
The globular clusters (GCs) of the Milky Way~(MW) are old gravitationally bound stellar systems with typical ages older than 6~Gyr and masses $\gtrsim$10$^{4}\rm M_{\odot}$~\cite{Kharchenko2013}. These objects can be used as a powerful tool to examine the Galactic structure and assembly the history at different scales from the formation of star clusters to hierarchical merger events~\cite{Kruijssen2020}. The recent high precision astrometric measurements provided by \textit{Gaia} Data Release~2~(DR2)~\cite{Gaia2018} allows us to calculate the mean proper motions for $\approx$150~GCs of the MW~\cite{Gaia2018,Baumgardt2019,Vasiliev2019}. In this work, by using two catalogues of GCs~\cite{Baumgardt2019,Vasiliev2019} containing the full 6D phase-space information, we performed the simulations of 148~GCs orbits aiming to test a possibility of the GCs collisions in the past 5~Gyr. Similar to previous studies, we study the dynamics of the GCs as the test-particle motion in axisymmetric MW-like potential~\cite{Gnedin1997, Pichardo2004, Allen2006, Allen2008, Moreno2014, Perez2018, Perez2020}.  

\begin{figure*}[htbp!]
\centering
\includegraphics[width=0.31\linewidth]{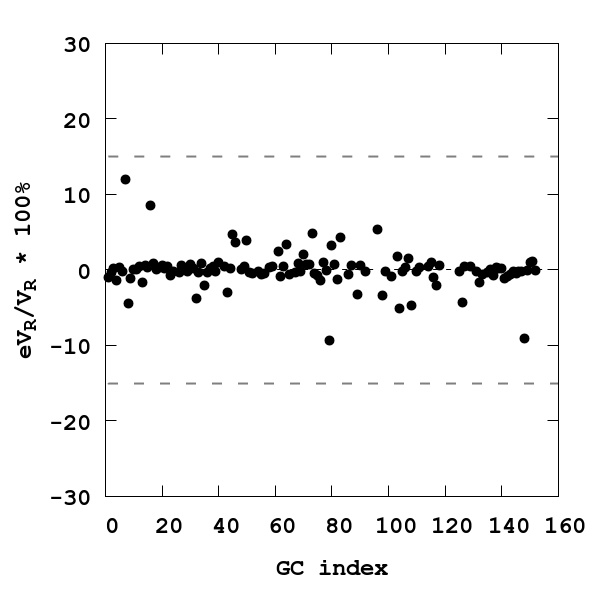} 
\includegraphics[width=0.31\linewidth]{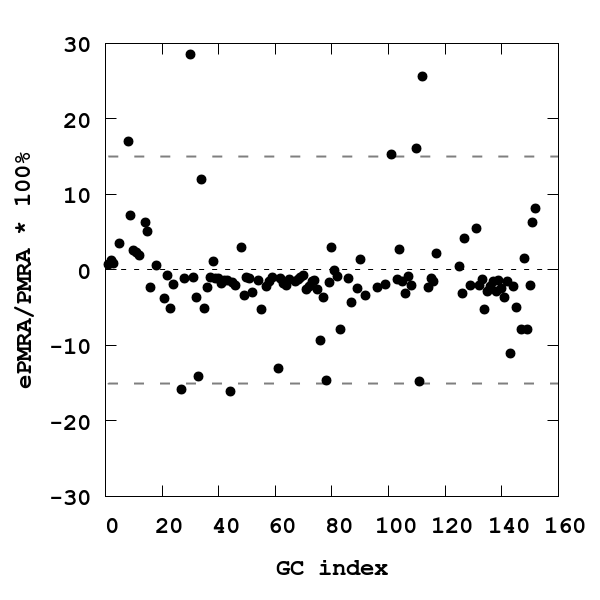} 
\includegraphics[width=0.31\linewidth]{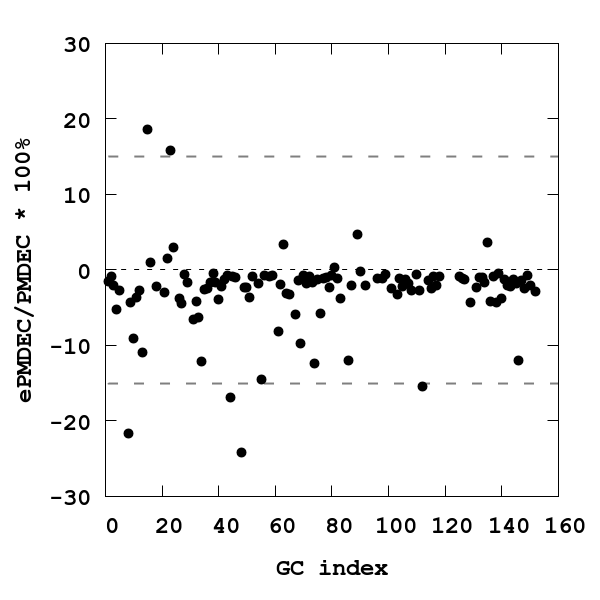} 
\caption{Distribution of the GCs measurement errors for radial velocity $v_{r}$~(left) and proper motions in right ascension~($\mu_{\alpha\ast}$, center) and in declination~($\mu_{\delta}$, right). Dashed grey horizontal lines indicate 15\% confidence range.}
\label{fig:PM}
\end{figure*}

\section*{\sc globular cluster sample}
\indent\indent 
Prior to the orbital integration, we prepared a complete catalogue of the MW GCs. That is, we merged two recent catalogues~\cite{Vasiliev2019,Baumgardt2019} which together contain the information about 152~objects (see Table~\ref{tab:GC3}). The resulting catalogue contains the complete phase-space information required for the initial conditions in our simulations: right ascension~(RA), declination~(DEC) and distance~(D), proper motions $\mu_{\alpha\ast}=\mu_{\alpha}\cos\delta$,  $\mu_{\delta}$ and radial velocity $v_{r}$

To avoid the calculation of the GCs orbits with large uncertainties in initial conditions we have analysed the errors of the \textit{Gaia} measurement. In Fig.~\ref{fig:PM} we show the relative errors for the radial velocity and proper motions where each GC has its own index (see Table~\ref{tab:GC3}). Thanks to the precise \textit{Gaia} measurements the uncertainties for the radial velocity ($v_{r}$) are quite small (mostly below 15\%). However, as it is seen, for proper motions ($\mu_{\alpha\ast}$, $\mu_{\delta}$) the situation is different. Therefore, we discard from our catalogue the GCs with the relative error larger than 30\% for radial velocity and proper motions. We found that only 8~GCs do not satisfy our selection and in Table~\ref{tab:GC3} these objects are marked with \textit{me} (measurement error).

For calculating positions and velocities in the Galactocentric rest-frame (for basic coordinate transformation see \cite{JS1987}), we assumed an in-plane distance of the Sun from the Galactic centre of $X_{_{\odot}}$=8.178~kpc~\cite{Gravity2019} and $Z_{_{\odot}}$=20.8~pc, a velocity of the Local Standard of Rest~(LSR), $V_{\rm LSR}$=234.737~km~s$^{-1}$~\cite{Mard2020}, and a peculiar velocity of the Sun with respect to the LSR, $U_\odot$=11.1~km~s$^{-1}$, $V_\odot$=12.24~km~s$^{-1}$, $W_\odot$=7.25~km~s$^{-1}$~\cite{Schonrich2010}.

\section*{\sc orbits integration}
\indent \indent 
For the GCs orbit integration we adopted the MW-like gravitational potential based on the superposition of \textit{bulge + disk + halo} models. In particular, the total potential consisting in a spherical bulge $\Phi_{\rm b}(R,z)$, an axisymmetric disk $\Phi_{\rm d}(R,z)$ and a spherical dark-matter halo $\Phi_{\rm h}(R,z)$ can be written as follows:
\begin{equation}
\Phi(R,z) = \Phi_{\rm b}(R,z)+ \Phi_{\rm d}(R,z)+ \Phi_{\rm h}(R,z)\,,
\end{equation}
where $R^{2}=x^{2}+y^{2}$ is the Galactocentric distance in polar coordinates and $z$ is the vertical coordinate perpendicular to the disk plane.

Potentials of the bulge and the disk were taken in the form of Miyamoto-Nagai~\cite{Miyamoto1975}, while the dark matter potential is assumed to be Navarro-Frenk-White (NFW)~\cite{Navarro1997}:
\begin{numcases}{}
\Phi_{\rm b}(R,z)=-\frac{M_{\rm b}}{(r^{2}+b_{\rm b}^{2})^{1/2}},\\
\Phi_{\rm d}(R,z)=-\frac{M_{\rm d}}{\left[R^{2}+\left(a_{\rm d}+\sqrt{z^{2}+b_{\rm d}^{2}}\right)^{2}\right]^{1/2}},\\
\Phi_{\rm h}(R,z)=-\frac{M_{\rm h}}{r}\ln\left(1+\frac{r}{b_{\rm h}}\right)\,,
\end{numcases}
where $\displaystyle r=\sqrt{R^2+z^2}$ is the spherical galactocentric distance. The values of masses and the scaling parameters of components can be found in Table~\ref{tab:pot-param} \cite{Bajkova2021, BajkovaAAT2021}.

For the GCs orbital integration we used a high-order parallel dynamical $N$-body code $\varphi$-GRAPE which is based on the fourth-order Hermite integration scheme with hierarchical individual block time steps scheme~\cite{Berczik2011}. More details about the code architecture and special GRAPE hardware can be found in~\cite{Harfst2007}. 

\begin{minipage}[l]{0.40\textwidth}
\begin{center}
\captionof{table}{Galactic potential parameters.}
\small{\begin{tabular}{lcc}
\hline
\hline
Parameter & Value & Unit \\
\hline
\hline
Bulge mass $M_{\rm b}$  & $1.03\times10^{10}$    & M$_{\odot}$ \\
Disk mass $M_{\rm d}$   & $6.51\times10^{10}$    & M$_{\odot}$  \\
Halo mass $M_{\rm h}$   & $29.00\times10^{10}$   & M$_{\odot}$ \\
Bulge scale param. $b_{\rm b}$   & $0.2672$      & kpc       \\
Disk scale param.  $a_{\rm d}$   & $4.4$         & kpc \\
Disk scale param.  $b_{\rm d}$   & $0.3084$      & kpc  \\
Halo scale param.  $b_{\rm h}$   & $7.7$         & kpc  \\
\hline
\end{tabular}}
\label{tab:pot-param}
\end{center}
\end{minipage}

\begin{figure*}[]
\centering
\includegraphics[width=0.32\linewidth]{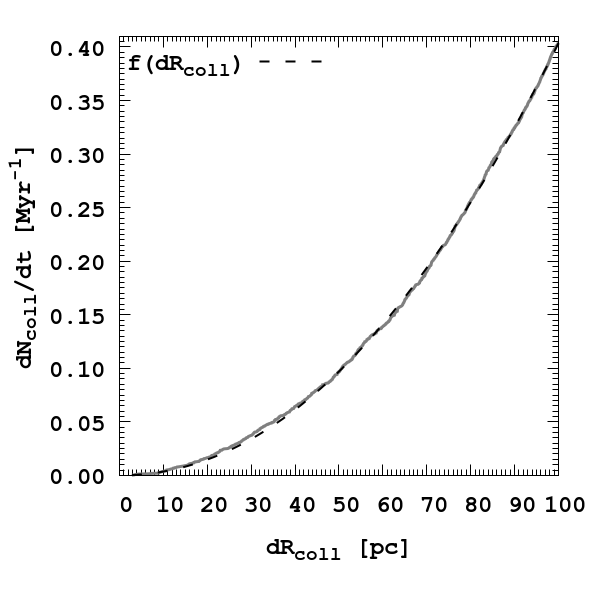} 
\includegraphics[width=0.32\linewidth]{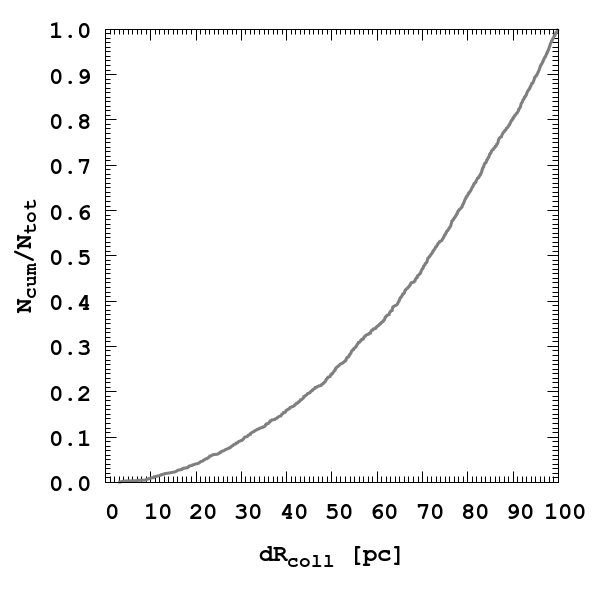}
\includegraphics[width=0.32\linewidth]{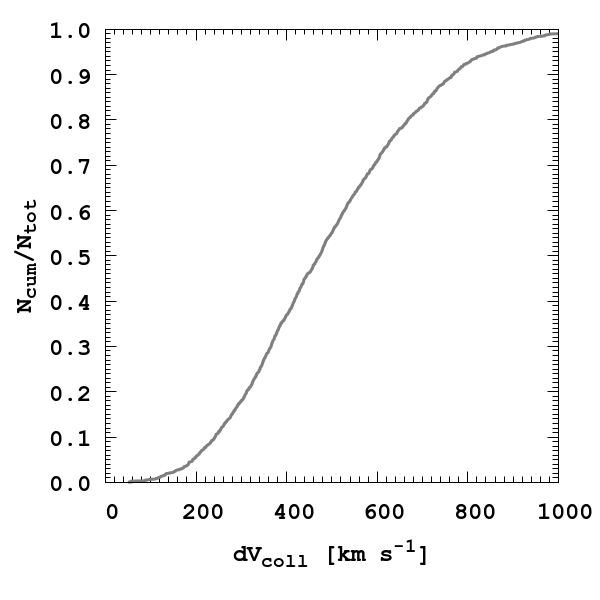} 
\caption{GCs collision rate as a function of the relative distance~(left), where black dashed line is a power-law fit~(see equation~\ref{eq:fit}). The normalized number of collisions as a function of the relative distance and relative velocity are shown in center and right panels, respectively.}
\label{fig:stat}
\end{figure*}

\newpage
\section*{\sc gc collision pairs}  
\indent \indent 
Before moving forward in the analysis of the collisions of the GCs population we have tested our numerical setup in order to keep tracking the GCs which orbits are the same during backward and forward integration. First, we integrated all 152~GCs backwards for 5~Gyr then we use the positions of velocities of all the GCs at the end of the simulations and integrate them forward for 5~Gyr. One could expect that the resulting positions and velocities should be identical to the observed ones. However, we have found that the orbits of 25~GCs are not invertible. These GCs usually pass by very close to the galactic center and most likely even an adaptive time-step is not able to capture their motions in the very center. Another possibility is a non-integrability of the potential which is hard to quantify and we leave this issue for further studies. Therefore, our final sample consists of 119~objects \cite{Ira2021}.

In order to count the number of collisions between pairs of GCs we used a set of three criteria. At the same time (i) a minimum separation between the GCs $dR_{\rm coll}$ should be $<$100~pc, (ii) the distance between the GCs should be less as twice of the sum of half-mass radii: $dR_{\rm coll} < 2(R_{{\rm hm},i}+R_{{\rm hm},j})$ and (iii) the relative velocity between objects $dV_{\rm coll}$ should be: $<$200~km~s$^{-1}$. 

According to the first criteria we have 2019 and 1973 collisions during backward and forward orbits integration, respectively. The second condition reduces these numbers to 38 collisions. Finally, applying the last condition we obtained only five reliable collision events. In Table~\ref{tab:gc-param} we show the characteristics of GC collisional pairs for reliable collisions (Terzan~3 -- NGC~6553, Terzan~3 -- NGC~6218, Liller~1 -- NGC~6522, Djorg~2 -- NGC~6553, NGC~6355 -- NGC~6637). It is worth mentioning, that all the colliding GCs were likely formed in the MW disk~\cite{Kruijssen2020}. 

\begin{minipage}[c]{0.40\textwidth}
\begin{center}
\captionof{table}{Characteristics of GC collisional pairs.}
\small{\begin{tabular}{llrcr}
\hline
\hline
\multicolumn{1}{c}{GC~1} & \multicolumn{1}{c}{GC~2} & \multicolumn{1}{c}{$dR_{\rm coll}$} & \multicolumn{1}{c}{$dV_{\rm coll}$} & \multicolumn{1}{c}{Time}  \\
    &      & \multicolumn{1}{c}{(pc)}          & \multicolumn{1}{c}{(km~s$^{-1}$)} & \multicolumn{1}{c}{(Myr)} \\
\hline
\hline
Terzan~3  & NGC~6553 & 25.58 & 148.18 & $237$\\
Terzan~3  & NGC~6218 & 10.75 & 183.12 & $581$\\
Liller~1  & NGC~6522 &  9.38 & 185.04 & $2625$\\
Djorg~2   & NGC~6553 & 20.22 & 153.14 & $2890$\\
NGC~6355  & NGC~6637 & 11.10 & 184.17 & 4886\\
\hline
\end{tabular}}
\label{tab:gc-param}
\end{center}
\end{minipage}

In order to estimate the global collision rate, in Fig.~\ref{fig:stat}~(left) we show the number of collisions per Myr as a function of impact parameter $dR_{\rm coll}$. The distribution can be well fitted by a simple power-law function:
\begin{equation}
\frac{dN_{\rm coll}}{dt}=10^{{\rm a}\cdot\lg(dR_{\rm coll})+{\rm b}},
\label{eq:fit}
\end{equation}
where ${\rm a}=2.057\pm0.001$ and ${\rm b}=-4.508\pm0.003$ are the fitting slope parameters. Therefore, we conclude that in each ten million years there is at least one collision with the impact parameter less than 50~pc.

In Fig.~\ref{fig:stat} we also present the normalized cumulative collisions number as a function of GC minimum impact parameter (center) and relative velocity at the moment of collision (right). As we can see the cumulative collision numbers can be also described by a power-law function, where the minimum values are $dR_{\rm coll}\approx3$~pc and $dV_{\rm coll}\approx85$~km~s$^{-1}$.

\section*{\sc interaction rates with central supermassive black hole}  

\begin{figure*}[]
\centering
\includegraphics[width=0.99\linewidth]{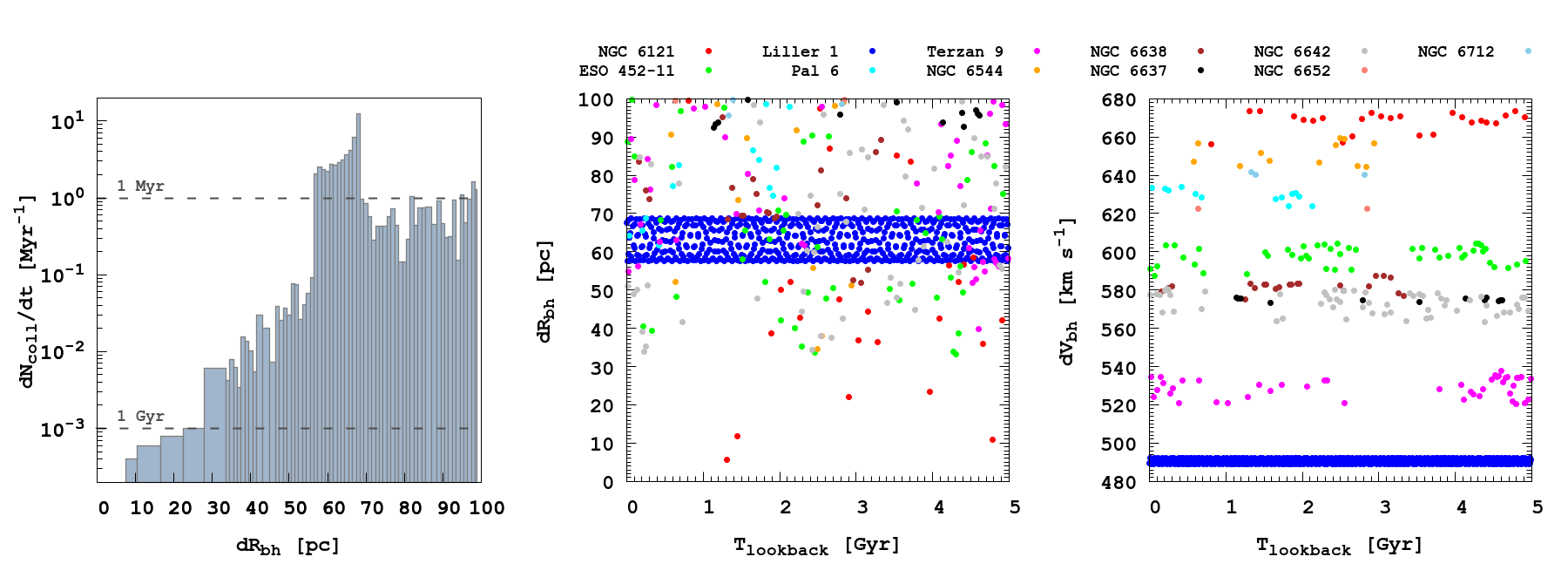}
\caption{Interaction rate of GCs with central SMBH~(left), where grey dashed lines are levels of the one event per Gyr and one event per Myr. The impact parameter from the center (middle) and the orbital velocity (right) of the GCs are shown.}
\label{fig:stat-bh}
\end{figure*}

\indent \indent 
In order to count a number of interactions with a central Supermassive Black Hole (SMBH) of GCs, we used criteria of minimum separation between the GC and central SMBH $dR_{\rm bh}$ should be $<100$~pc. In order to estimate the interaction rate of GC with central SMBH, in Fig.~\ref{fig:stat-bh}~(left) we show the number of events rate per Myr as a function of  GC impact parameter $dR_{\rm bh}$. According to this figure we can estimate the close interaction rate as one event per Gyr with the impact parameter less than 30 pc. Also we can conclude that we have at least one event per Myr with the impact parameter less than 60~pc. 

According to the above criteria we can estimate 11 very close encounter events: NGC~6121, ESO~452-11, Liller~1, Pal~6, Terzan~9, NGC~6544, NGC~6638, NGC~6637, NGC~6642, NGC~6652 and NGC~6712. All of them have a very close passing orbit trajectory and a high probability of interaction with Milky Way SMBH. In  Fig.~\ref{fig:stat-bh} (center and right panels) we show the impact parameter from the center and the orbital velocity of the actual GCs. Each of the above 11 events is marked by different colors. As we can see from these panels the GC Liller~1 during the last five Gyr always have a close pericenter passage at a level of around 65~pc (blue points). The closest encounter with the central SMBH in our simulation has a 5.5~pc. This is the NGC~6121 GC (red points). The relatively high velocities can be easily explained by the strong dynamical influence of the SMBH on the orbital motion of GCs.

In Table~\ref{tab:bh-param} we present the interaction events of GCs with minimum separation (second column) from SMBH. On the third column we show the corresponding pericenter passage velocity and the time past when this event happens (fourth column). After the extended literature search we try to identify the GCs possible progenitors. In most cases the clusters have an MW main bulge origin. 

\hspace{-1.5em}
\begin{minipage}[c]{0.48\textwidth}
\begin{center}
\captionof{table}{Characteristics of GCs that have closing pass with the central supermassive black hole.}
\small{\begin{tabular}{lccrl}
\hline
\hline
\multicolumn{1}{c}{GC} & \multicolumn{1}{c}{$dR_{\rm bh, min}$} & \multicolumn{1}{c}{$dV_{\rm bh}$} & \multicolumn{1}{c}{Time}  & Progenitor \\
    & \multicolumn{1}{c}{(pc)}          & \multicolumn{1}{c}{(km~s$^{-1}$)} & \multicolumn{1}{c}{(Myr)} & \\
\hline
\hline
NGC~6121    & 5.5 & 673 & 1314 & Kraken \cite{Kruijssen2020}\\
ESO~452-11  & 33  & 604 & 4303 & -- \\
Liller~1    & 58  & 499 & 1859 & XXX \cite{Massari2019}\\
            &     &     &      & MW \cite{Bastian2021}\\
Pal~6       & 62  & 634 & 427  & LE \cite{Massari2019}\\
            &     &     &      & MB \cite{Souza2021}\\
Terzan~9    & 40  & 637 & 4599 & MB \cite{Massari2019}\\
NGC~6544    & 35  & 569 & 2499 & Kraken \cite{Kruijssen2020}\\
NGC~6638    & 52  & 587 & 3060 & MB \cite{Massari2019}\\
NGC~6637    & 92  & 575 & 1136 & MB \cite{Massari2019}\\
NGC~6642    & 34  & 580 & 2243 & MB \cite{Massari2019}\\
NGC~6652    & 99  & 622 & 640  & MB \cite{Massari2019}\\
NGC~6712    & 96  & 641 & 1330 & LE \cite{Massari2019}\\
            &     &     &      & Kraken \cite{Kruijssen2020}\\
\hline
\multicolumn{5}{p{.95\textwidth}}{NOTE: 
Column Progenitor contains possible GC's origin with reference: Kraken - GC stands to Kraken accretion event, XXX - GC does not have available kinematics, LE - GC stands to the unassociated low-energy group, MB - GC stands to the main bulge, MW - GC stands to the \textit{in situ} formation.} \\
\end{tabular}}
\label{tab:bh-param}
\end{center}
\end{minipage}

\begin{figure}[htbp!]
\centering
\includegraphics[width=0.98\linewidth]{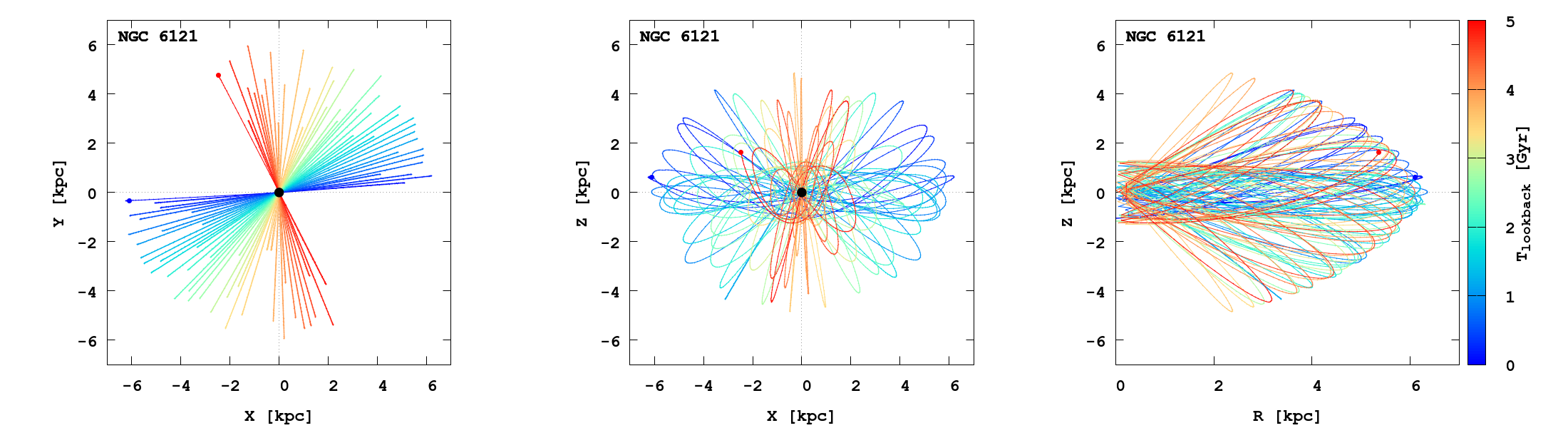}
\includegraphics[width=0.98\linewidth]{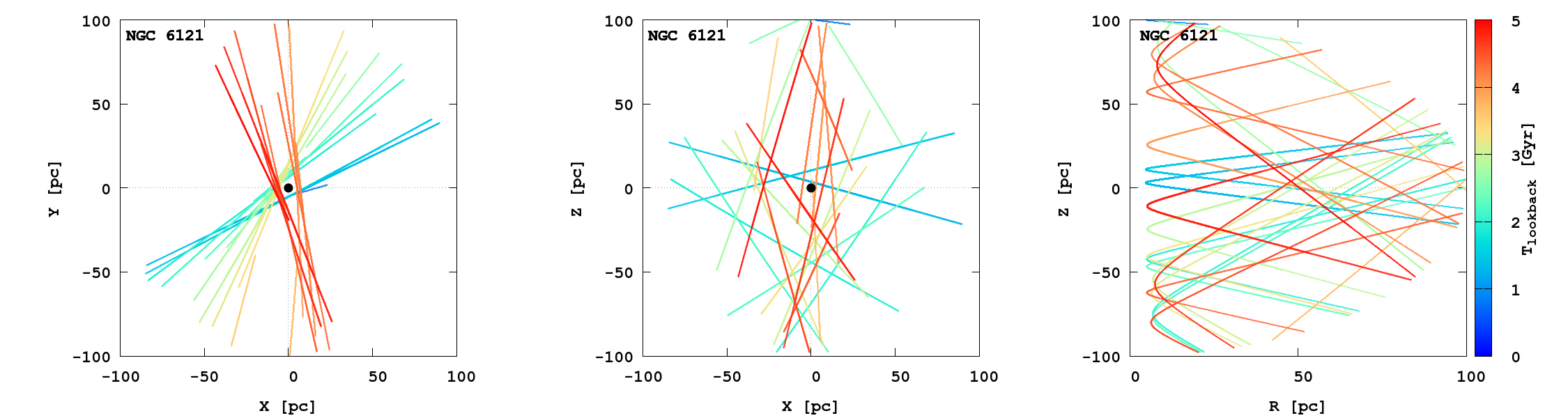}
\includegraphics[width=0.98\linewidth]{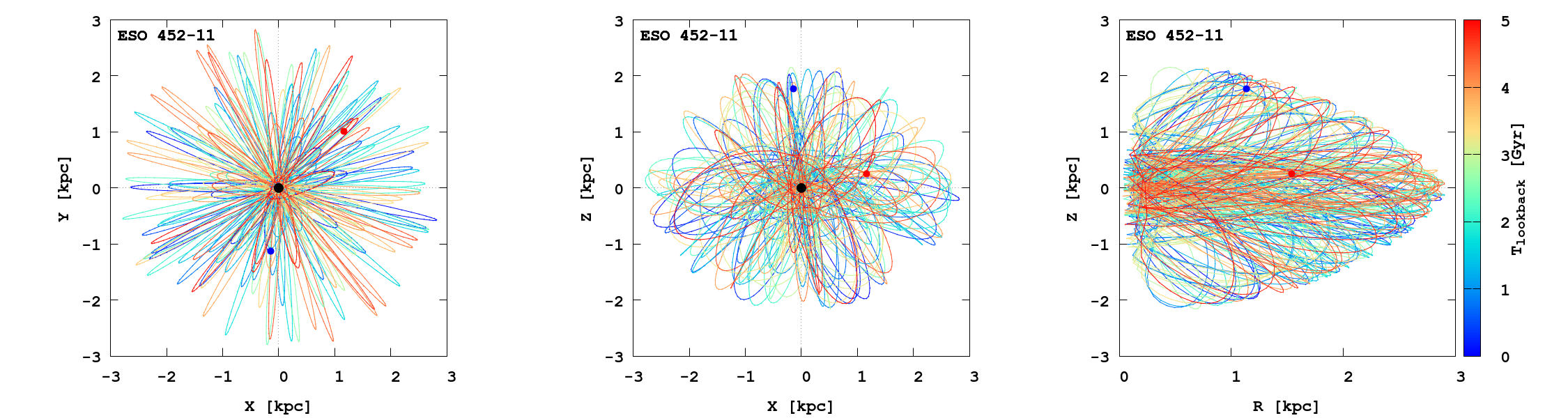}
\includegraphics[width=0.98\linewidth]{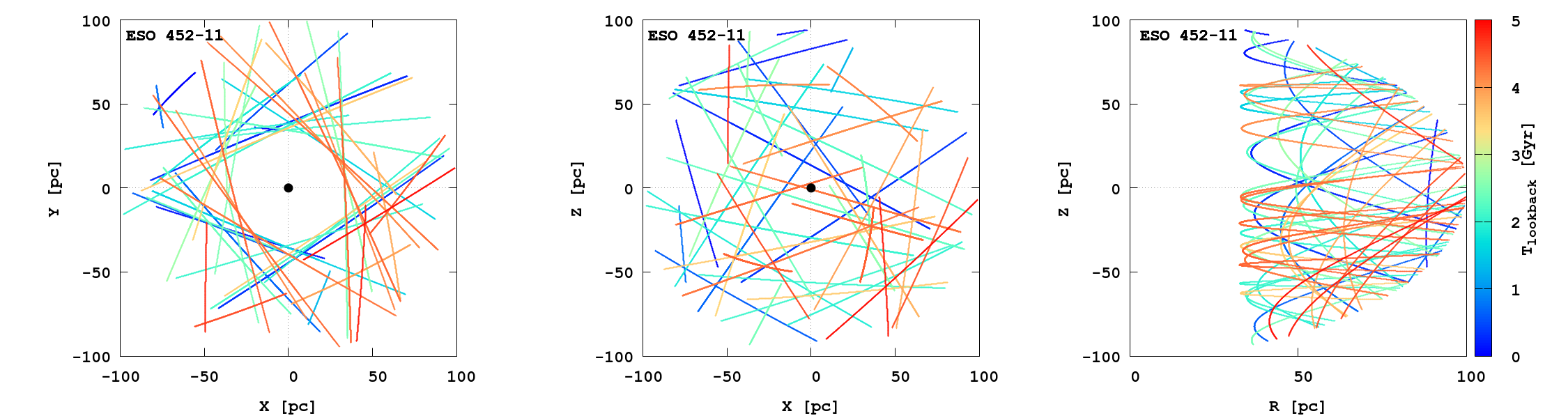}
\caption{GC NGC~6121 and ESO~452-11 (from top to bottom) orbits in $X-Y$ plane (left), in $X-Z$ plane (middle) and in $Z-R$ plane (right), where $R$ is distance in the Galactic plane.}
\label{fig:orb1}
\end{figure}

\begin{figure}[htbp!]
\centering
\includegraphics[width=0.98\linewidth]{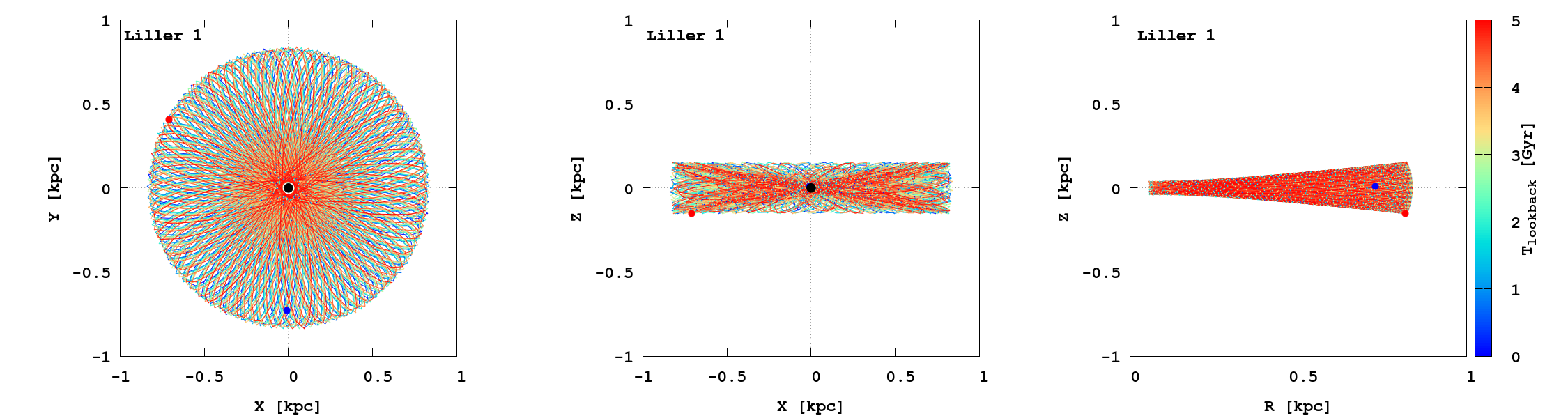}
\includegraphics[width=0.98\linewidth]{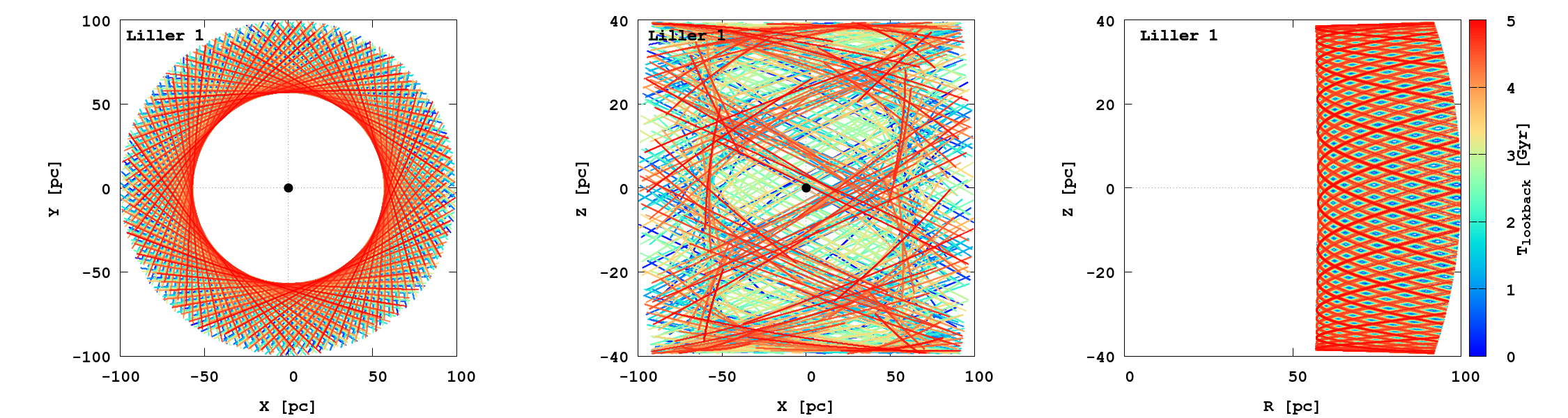}
\includegraphics[width=0.98\linewidth]{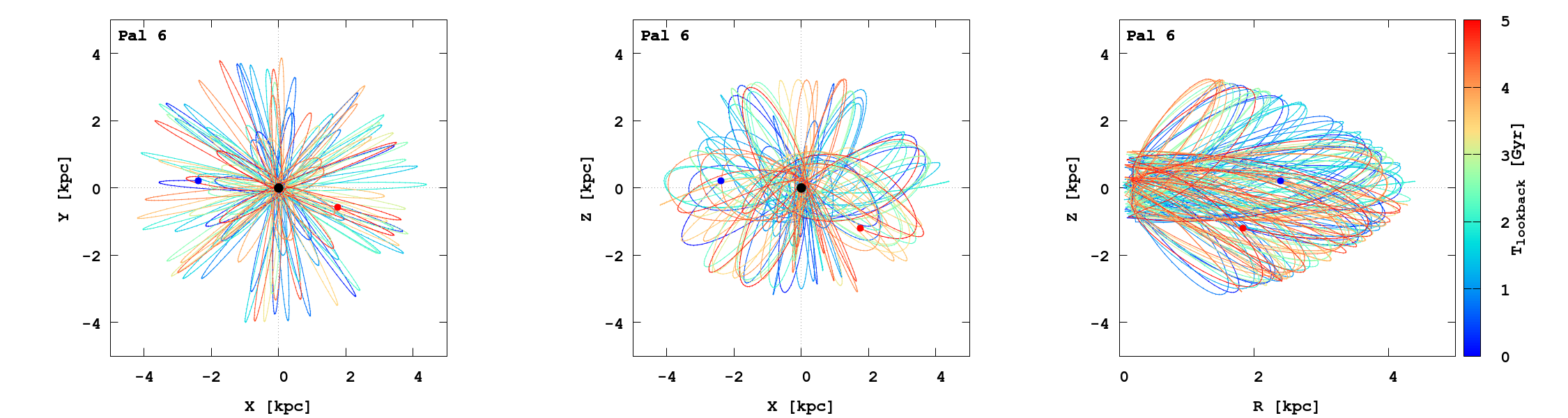}
\includegraphics[width=0.98\linewidth]{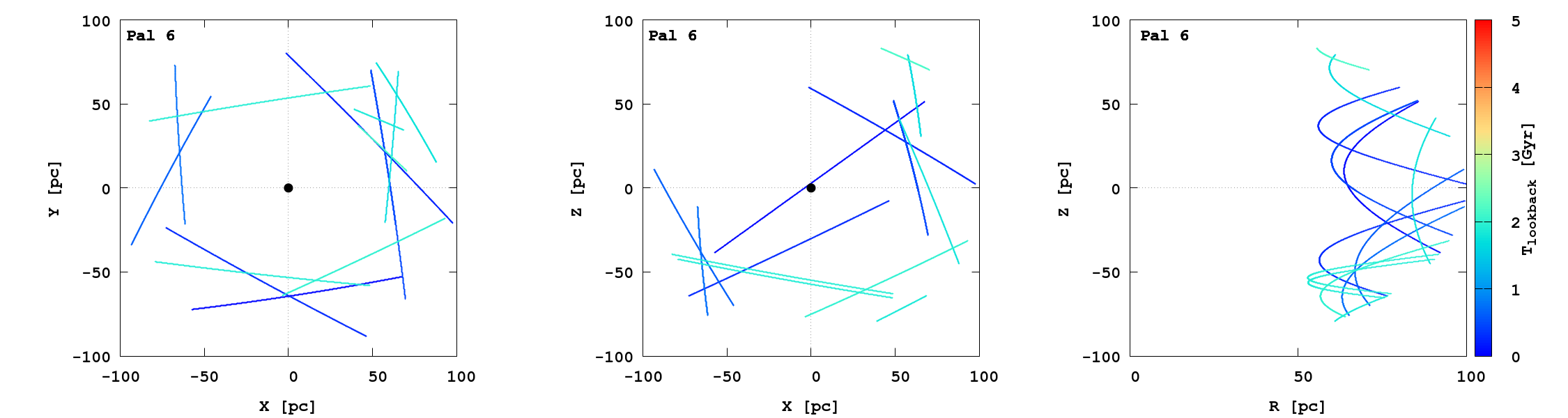}
\caption{As in Fig.~\ref{fig:orb1} for Liller~1 and Pal~6.}
\label{fig:orb2}
\end{figure}

\begin{figure}[htbp!]
\centering
\includegraphics[width=0.98\linewidth]{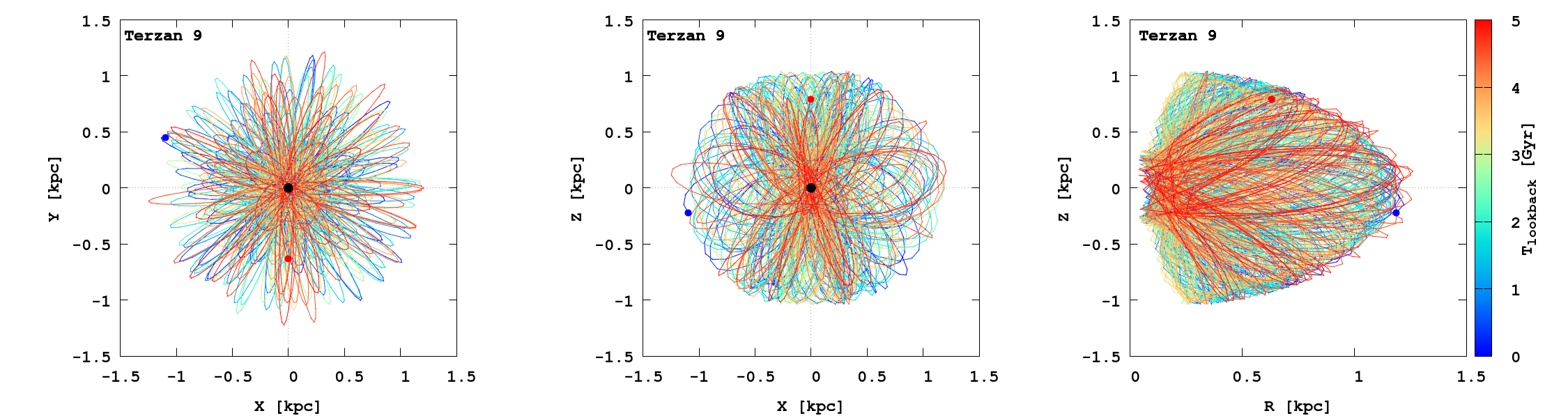}
\includegraphics[width=0.98\linewidth]{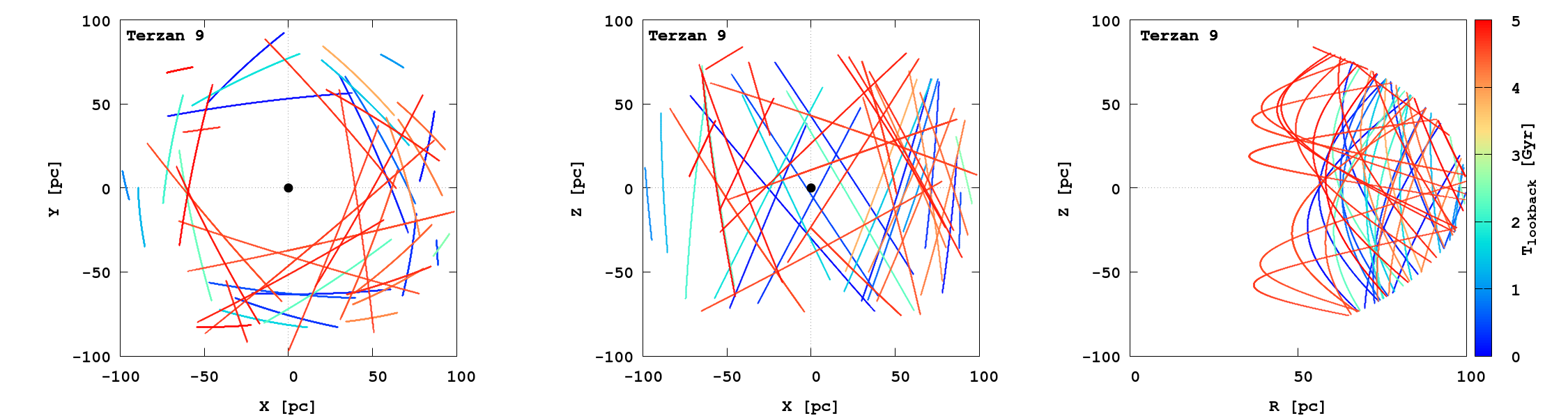}
\includegraphics[width=0.98\linewidth]{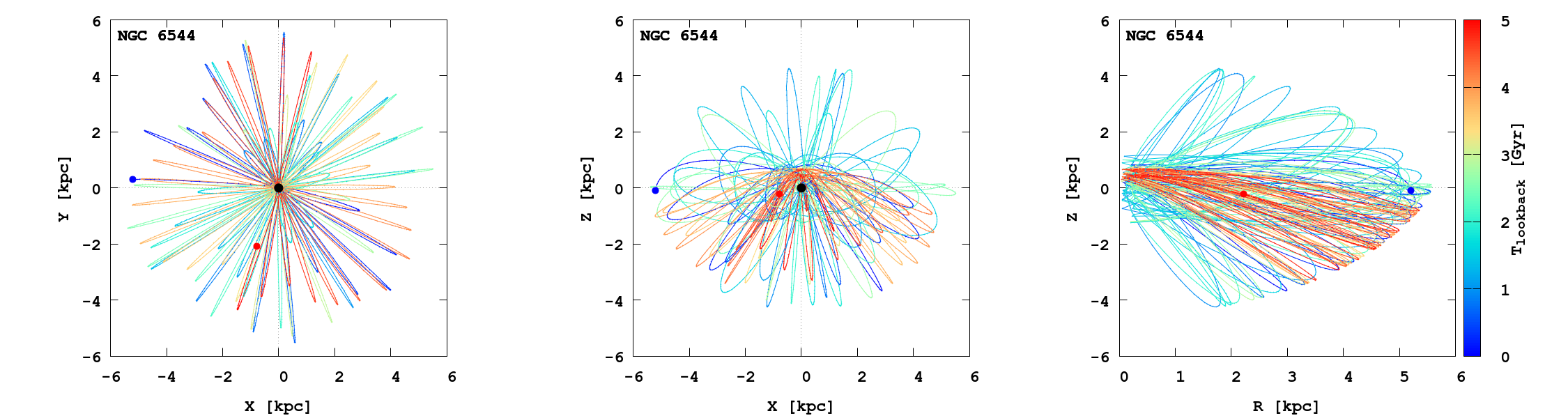}
\includegraphics[width=0.98\linewidth]{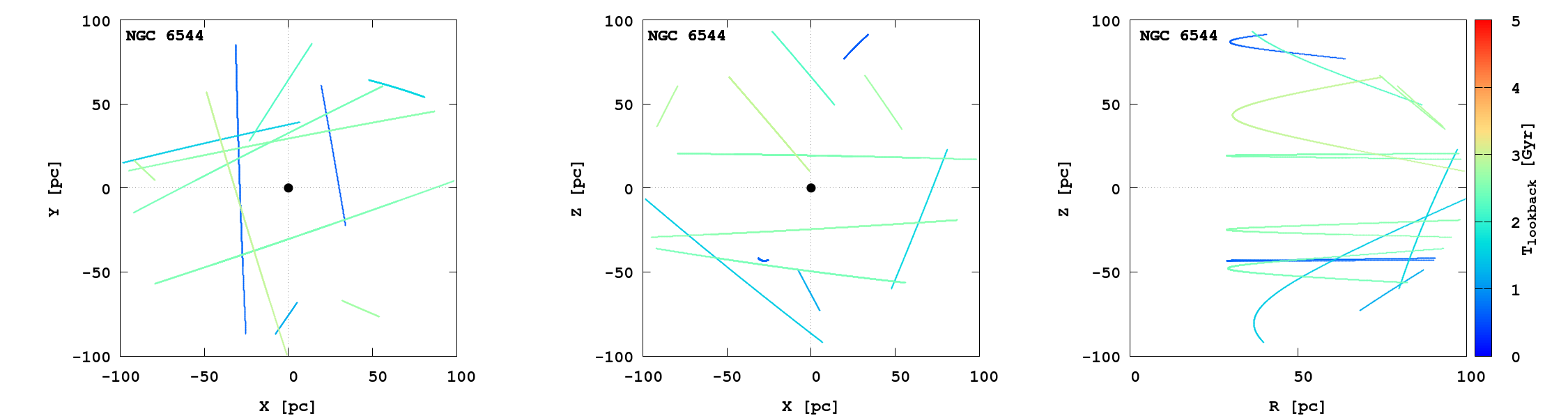}
\caption{As in Fig.~\ref{fig:orb1} for Terzan~9 and NGC~6544.}
\label{fig:orb3}
\end{figure}

\begin{figure}[htbp!]
\centering
\includegraphics[width=0.98\linewidth]{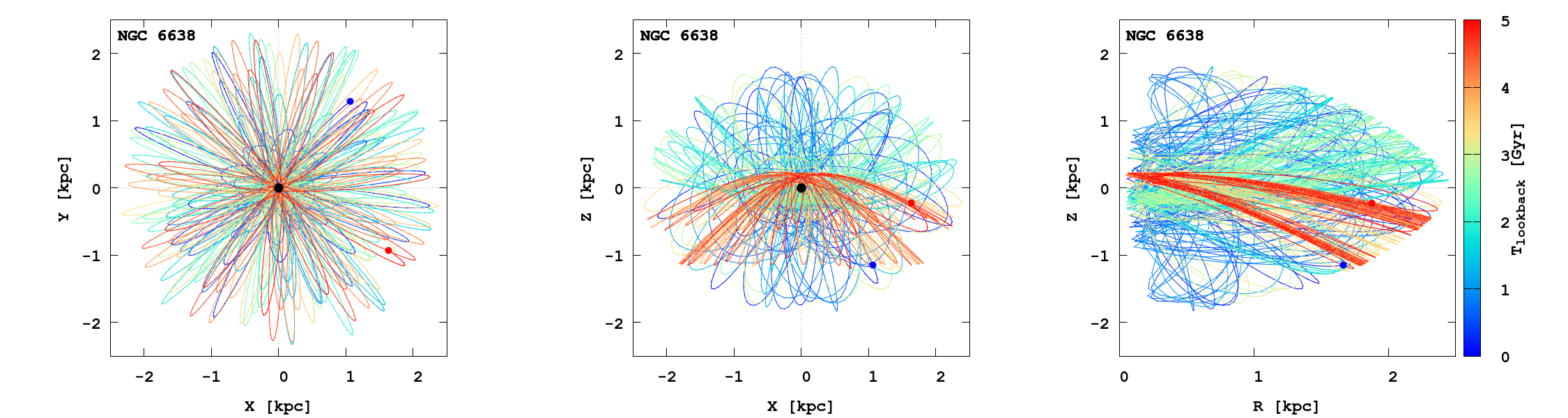}
\includegraphics[width=0.98\linewidth]{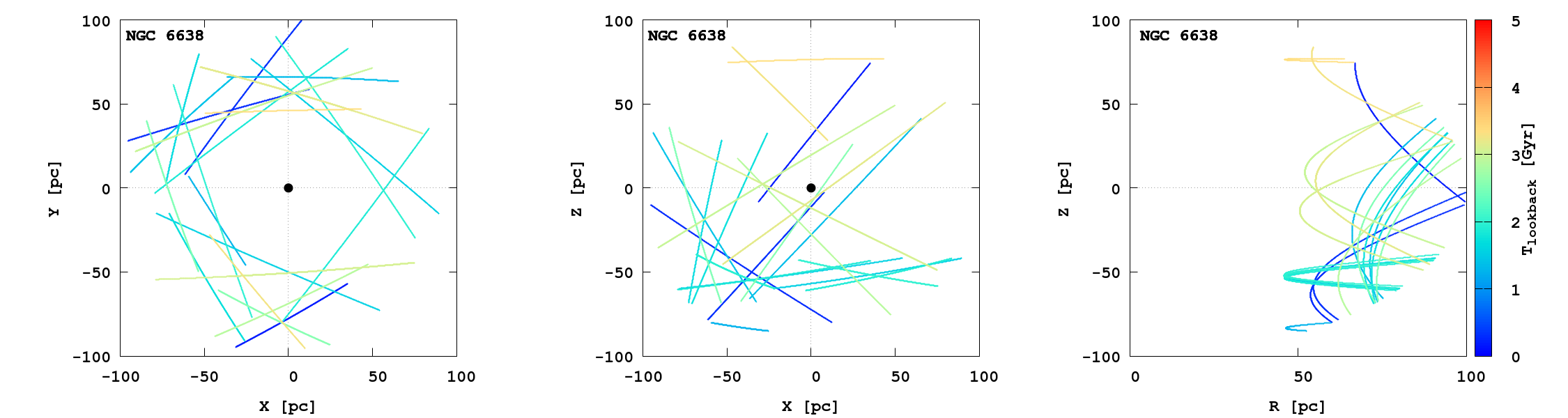}
\includegraphics[width=0.98\linewidth]{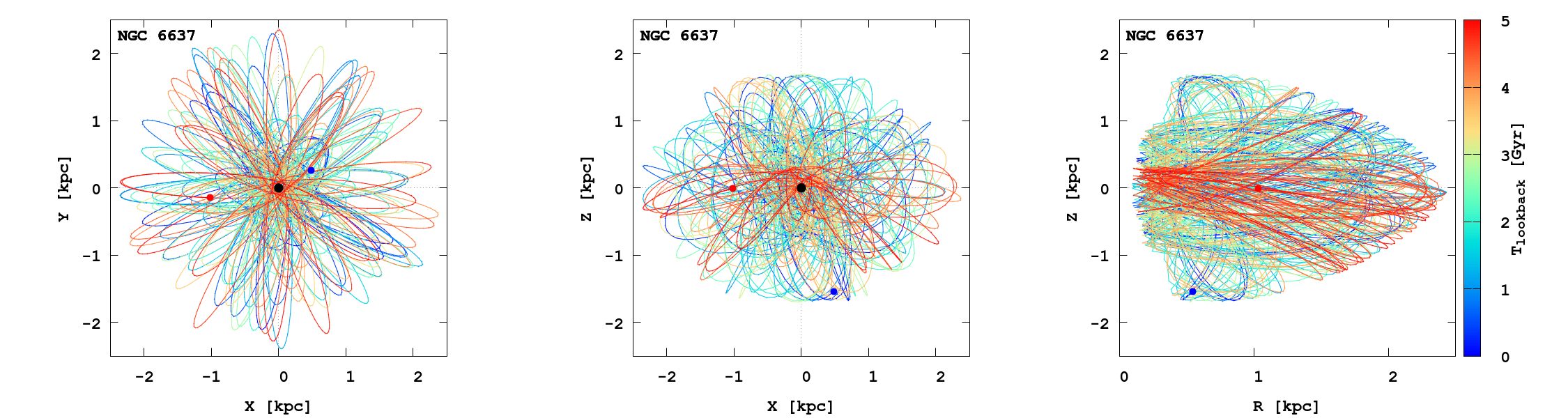}
\includegraphics[width=0.98\linewidth]{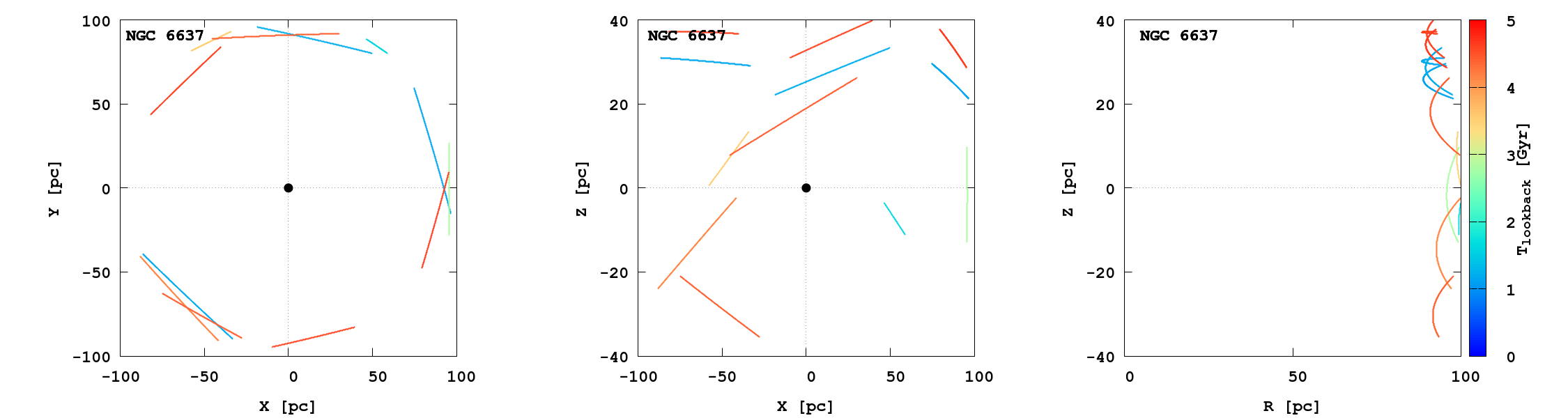}
\caption{As in Fig.~\ref{fig:orb1} for NGC~6638 and NGC~6637.}
\label{fig:orb4}
\end{figure}

\begin{figure}[htbp!]
\centering
\includegraphics[width=0.98\linewidth]{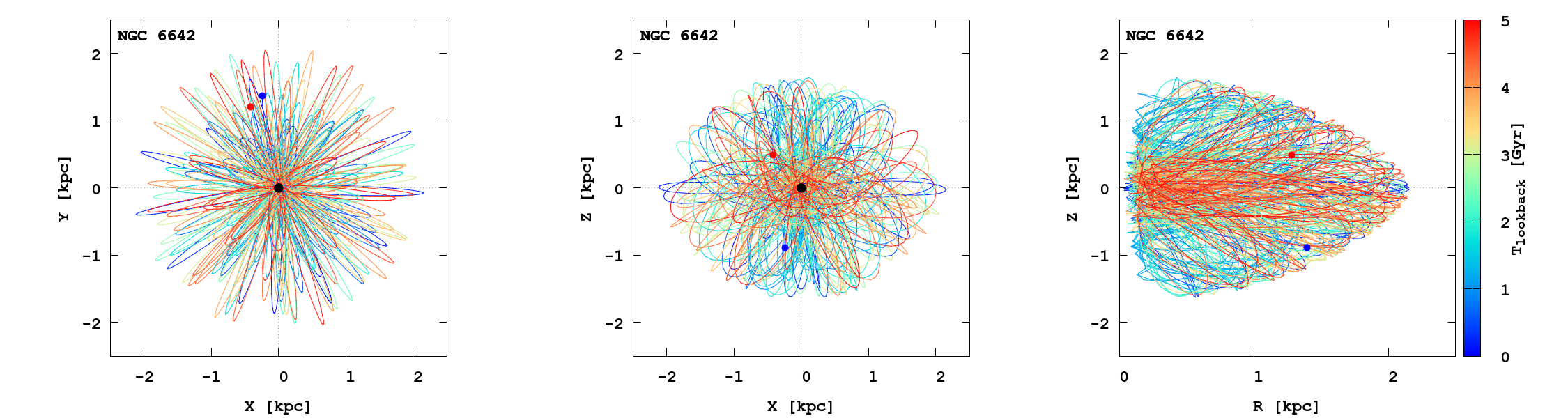}
\includegraphics[width=0.98\linewidth]{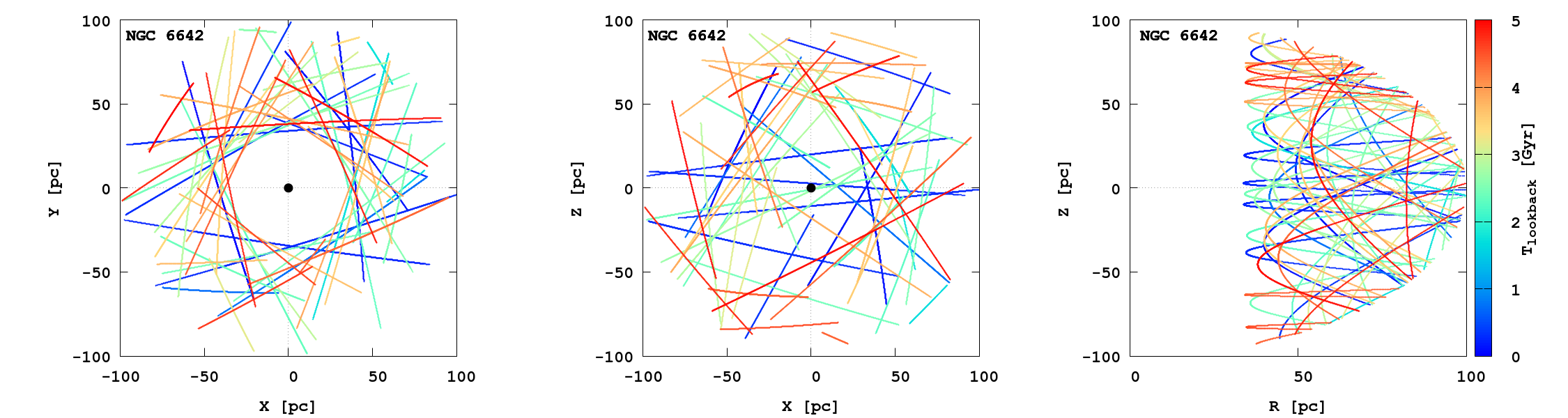}
\includegraphics[width=0.98\linewidth]{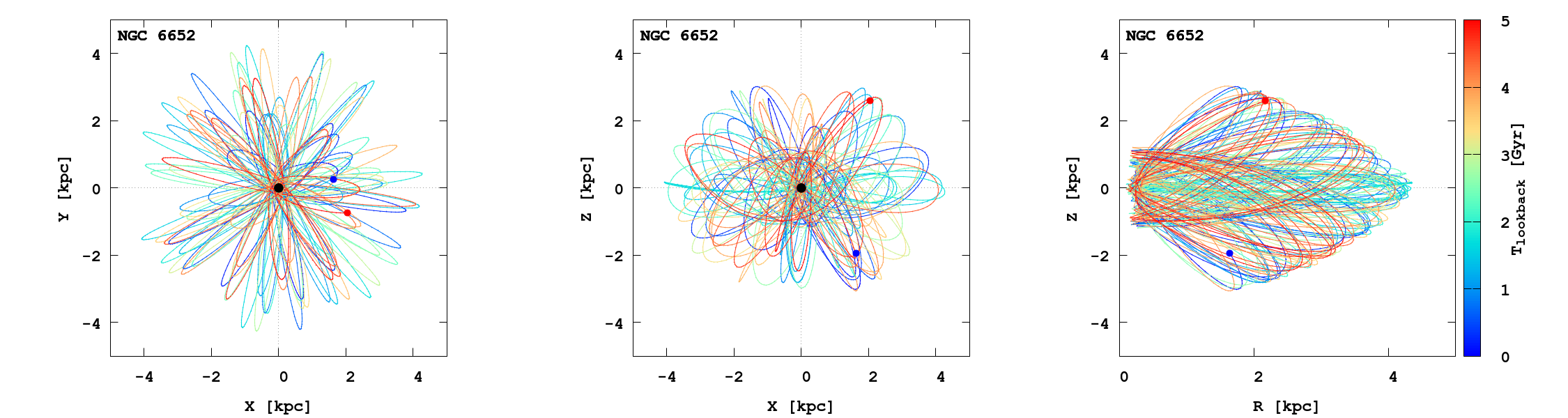}
\includegraphics[width=0.98\linewidth]{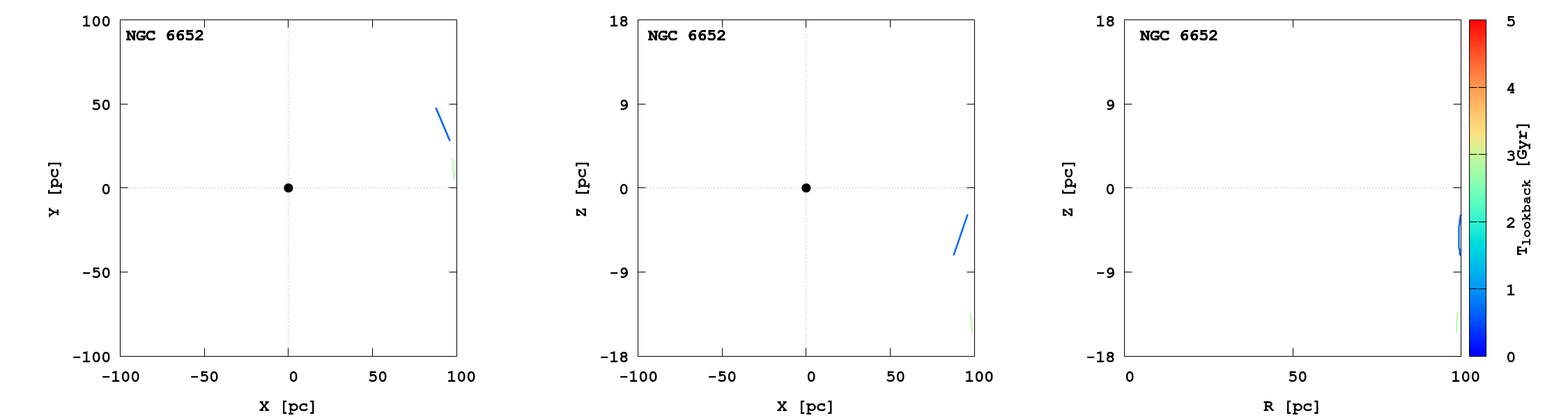}
\caption{As in Fig.~\ref{fig:orb1} for NGC~6642 and NGC~6652.}
\label{fig:orb5}
\end{figure}

\begin{figure}[htbp!]
\centering
\includegraphics[width=0.98\linewidth]{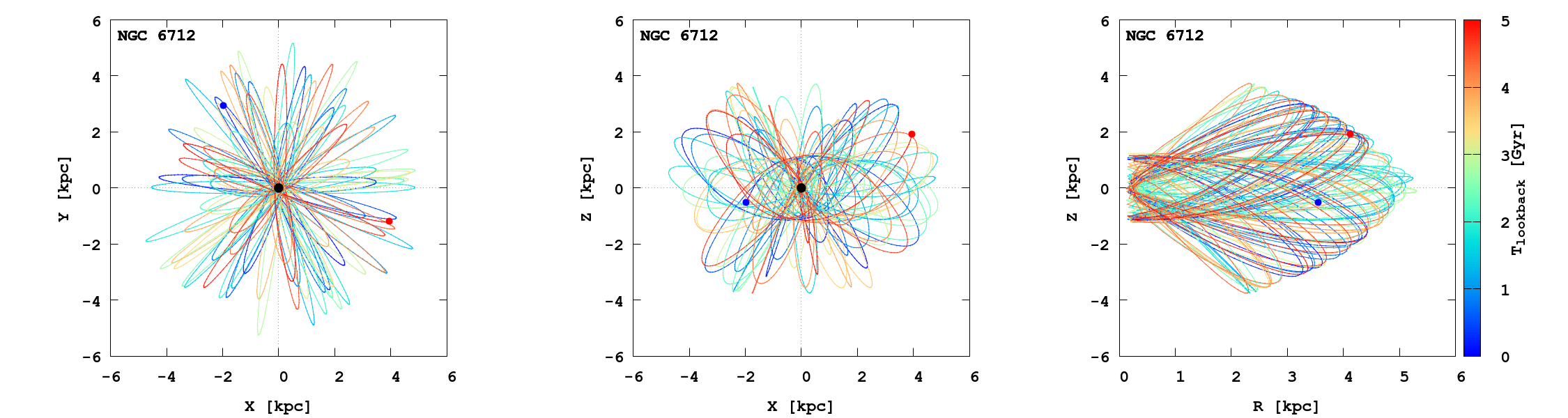}
\includegraphics[width=0.98\linewidth]{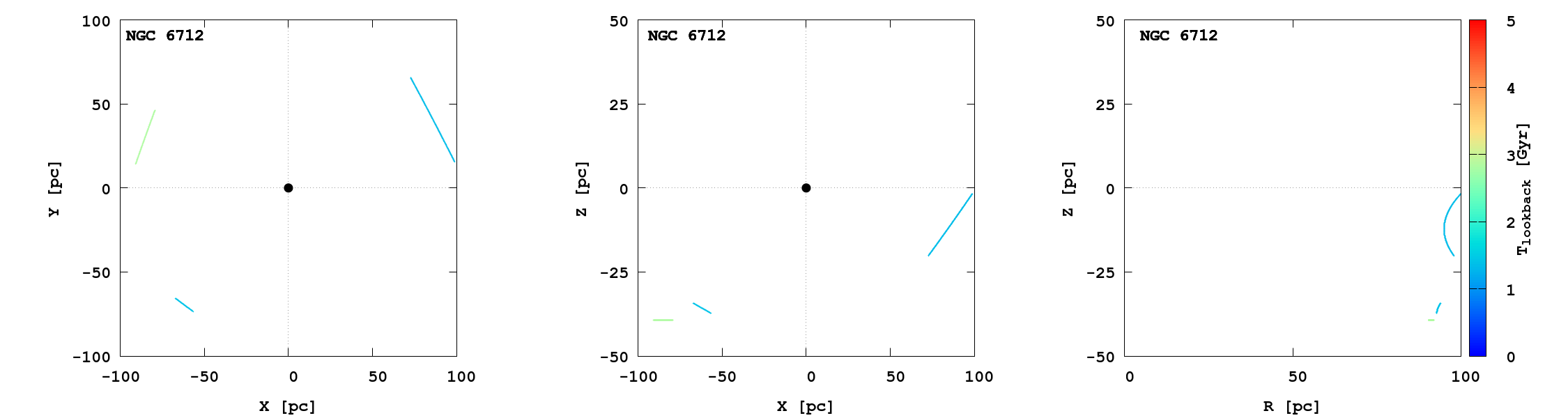}
\caption{As in Fig.~\ref{fig:orb1} for NGC~6712.}
\label{fig:orb6}
\end{figure}

The detail interaction of the selected 11~GCs with the central SMBH we present in Fig.~\ref{fig:orb1} - Fig.~\ref{fig:orb6}. The upper figures for each objects show`s the global view of object trajectories. The bottom figures show`s the detailed view of encounters. As we can see from the detailed visualization the GCs during the five Gyr of integration the GCs came to the central SMBH quite often. On the summary Table~\ref{tab:GC3} we present all the 152~globular clusters with full information about the clusters. As we can note the Liller~1 GC have both collisions with other cluster and also close interaction with central SMBH. 

\section*{\sc conclusions}
\indent \indent 
Using the present-day \textit{Gaia} DR~2-based catalogues~\cite{Vasiliev2019,Baumgardt2019} we have analyzed the orbits of the Milky Way globular clusters. From 152~GCs we discard 8~objects with large velocity errors. For the remaining 146~GCs, we analyse both backward and forward orbits calculated in the MW-like external potential using our developed high order $\varphi$-GRAPE code. Using complex criteria for the collisions detection we robustly identified five colliding pairs: Terzan~3 -- NGC~6553, Terzan~3 -- NGC~6218, Liller~1 -- NGC~6522, Djorg~2 -- NGC~6553, NGC~6355 -- NGC~6637. We also estimated the overall collision rate as about one collision with the impact parameter less than 50~pc per 10~Myr. 

Also we analyzed the GCs interaction rates with the central supermassive black hole. Assuming the maximum 100~pc distance criteria for separation between them we estimated 11 close encounter events: NGC~6121, ESO~452-11, Liller~1, Pal~6, Terzan~9, NGC~6544, NGC~6638, NGC~6637, NGC~6642, NGC~6652 and NGC~6712. From our numerical simulations we estimate the close interaction rate as: at least one event per Gyr with the impact parameter less than 30~pc; and one event per Myr with the impact parameter less than 60~pc. Our calculations show one very close encounter of NGC~6121 with the central SMBH near 5.5~pc (practically direct collision). Based on the extended literature search for the possible progenitor of our selected 11~GCs, we found that most of them have a Milky Way main bulge origin.
\section*{\sc acknowledgement}
\indent \indent 
The work of MI and MS was supported by the National Academy of Sciences of Ukraine under research project of young scientists No.~0121U111799. 
The work of MK and BS has been funded by the Science Committee of the Ministry of Education and Science of the Republic of Kazakhstan (Grant No. AP08856149, AP08856184 and BR10965141).
PB acknowledges support by the Chinese Academy of Sciences through the Silk Road Project at NAOC, the President’s 
International Fellowship (PIFI) for Visiting Scientists program of CAS.
The work of PB, MI and MS was also supported by the Volkswagen Foundation under the Trilateral Partnerships grants No.~90411 and~97778. 
This work has made use of data from the European Space Agency (ESA) mission GAIA (\url{https://www.cosmos.esa.int/gaia}), processed by the GAIA Data Processing and Analysis Consortium (DPAC, \url{https://www.cosmos.esa.int/web/gaia/dpac/consortium}). Funding for the DPAC has been provided by national institutions, in particular the institutions participating in the GAIA Multilateral Agreement. 

The authors is grateful to an anonymous referee for useful comments and suggestions that helped to improve the manuscript.

\end{multicols}

\begin{table}[h!]
\renewcommand{\arraystretch}{1.0}
\centering
\caption{Initial list of GCs.}
{\small
\begin{tabular}{l@{\hspace{0.95\tabcolsep}}l@{\hspace{0.95\tabcolsep}}c@{\hspace{0.95\tabcolsep}}|l@{\hspace{0.95\tabcolsep}}l@{\hspace{0.95\tabcolsep}}c@{\hspace{0.95\tabcolsep}}|l@{\hspace{0.95\tabcolsep}}l@{\hspace{0.95\tabcolsep}}c@{\hspace{0.95\tabcolsep}}|l@{\hspace{0.95\tabcolsep}}l@{\hspace{0.95\tabcolsep}}c@{\hspace{0.95\tabcolsep}}|l@{\hspace{0.95\tabcolsep}}l@{\hspace{0.95\tabcolsep}}c}
\hline
\hline
ID & Name & Flag & ID & Name & Flag & ID & Name & Flag & ID & Name & Flag & ID & Name & Flag \\
\hline
\hline
1  & NGC~104   &    & 32 & NGC~5634  &    & 63          & NGC~6273  &    & 94  & Terzan~5  &    & 125 & NGC~6656  &    \\
2  & NGC~288   &    & 33 & NGC~5694  &    & 64          & NGC~6284  &    & 95  & NGC~6440  &    & 126 & Pal~8     &    \\
3  & NGC~362   &    & 34 & IC~4499   &    & 65          & NGC~6287  &    & 96  & NGC~6441  &    & 127 & NGC~6681  &    \\
4  & Whiting~1 &    & 35 & NGC~5824  &    & 66          & NGC~6293  &    & 97  & Terzan~6  &    & 128 & NGC~6712  & bh \\
5  & NGC~1261  &    & 36 & Pal~5     &    & 67          & NGC~6304  &    & 98  & NGC~6453  &    & 129 & NGC~6715  &    \\
6  & Pal~1     & me & 37 & NGC~5897  &    & 68          & NGC~6316  &    & 99  & NGC~6496  &    & 130 & NGC~6717  &    \\
7  & E~1       & me & 38 & NGC~5904  &    & 69          & NGC~6341  &    & 100 & Terzan~9  & bh & 131 & NGC~6723  &    \\
8  & Eridanus  &    & 39 & NGC~5927  &    & 70          & NGC~6325  &    & 101 & Djorg~2   & cc & 132 & NGC~6749  &    \\
9  & Pal~2     &    & 40 & NGC~5946  &    & 71          & NGC~6333  &    & 102 & NGC~6517  &    & 133 & NGC~6752  &    \\
10 & NGC~1851  &    & 41 & BH~176    & me & 72          & NGC~6342  &    & 103 & Terzan~10 &    & 134 & NGC~6760  & me \\
11 & NGC~1904  &    & 42 & NGC~5986  &    & 73          & NGC~6356  &    & 104 & NGC~6522  & cc & 135 & NGC~6779  &    \\
12 & NGC~2298  &    & 43 & FSR~1716  &    & 74          & NGC~6355  & cc & 105 & NGC~6535  &    & 136 & Terzan~7  &    \\
13 & NGC~2419  &    & 44 & Pal~14    &    & 75          & NGC~6352  &    & 106 & NGC~6528  &    & 137 & Pal~10    &    \\
14 & Pyxis     &    & 45 & BH~184    &    & 76          & IC~1257   &    & 107 & NGC~6539  &    & 138 & Arp~2     &    \\
15 & NGC~2808  &    & 46 & NGC~6093  &    & 77          & Terzan~2  &    & 108 & NGC~6540  &    & 139 & NGC~6809  &    \\
16 & E~3       &    & 47 & NGC~6121  & bh & 78          & NGC~6366  &    & 109 & NGC~6544  & bh & 140 & Terzan~8  &    \\
17 & Pal~3     & me & 48 & NGC~6101  &    & 79          & Terzan~4  &    & 110 & NGC~6541  &    & 141 & Pal~11    &    \\
18 & NGC~3201  &    & 49 & NGC~6144  &    & 80          & BH~229    &    & 111 & ESO~280-6 &    & 142 & NGC~6838  &    \\
19 & Pal~4     & me & 50 & NGC~6139  &    & 81$^{\ast}$ & FSR~1758  &    & 112 & NGC~6553  & cc & 143 & NGC~6864  &    \\
20 & Crater    &    & 51 & Terzan~3  & cc & 82          & NGC~6362  &    & 113 & NGC~6558  &    & 144 & NGC~6934  &    \\
21 & NGC~4147  &    & 52 & NGC~6171  &    & 83$^{\ast}$ & Liller~1  & cc, bh & 114 & Pal~7     &    & 145 & NGC~6981  &    \\
22 & NGC~4372  &    & 53 & ESO~452-11 & bh & 84          & NGC~6380  &    & 115 & Terzan~12 &    & 146 & NGC~7006  &    \\
23 & Rup~106   &    & 54 & NGC~6205  &    & 85          & Terzan~1  &    & 116 & NGC~6569  &    & 147 & NGC~7078  &    \\
24 & NGC~4590  &    & 55 & NGC~6229  &    & 86          & Ton~2     &    & 117 & BH~261    &    & 148 & NGC~7089  &    \\
25 & NGC~4833  &    & 56 & NGC~6218  & cc & 87          & NGC~6388  &    & 118 & NGC~6584  &    & 149 & NGC~7099  &    \\
26 & NGC~5024  &    & 57 & FSR~1735  & me & 88          & NGC~6402  &    & 119 & NGC~6624  &    & 150 & Pal~12    &    \\
27 & NGC~5053  &    & 58 & NGC~6235  &    & 89          & NGC~6401  &    & 120 & NGC~6626  &    & 151 & Pal~13    &    \\
28 & NGC~5139  &    & 59 & NGC~6254  &    & 90          & NGC~6397  &    & 121 & NGC~6638  & bh & 152 & NGC~7492  &    \\
29 & NGC~5272  &    & 60 & NGC~6256  &    & 91          & Pal~6     & bh & 122 & NGC~6637  & cc, bh &     &           &    \\
30 & NGC~5286  & me & 61 & Pal~15    &    & 92          & NGC~6426  &    & 123 & NGC~6642  & bh &     &           &    \\
31 & NGC~5466  &    & 62 & NGC~6266  &    & 93          & Djorg~1   &    & 124 & NGC~6652  & bh &     &           &    \\
\hline
\multicolumn{15}{p{.95\textwidth}}{NOTE: Parameters for all GCs was taken from~\cite{Vasiliev2019} with the exception for GCs marked~$^{\ast}$ with data from~\cite{Baumgardt2019}. Column Flag contains additional information: me - GC excluded from the integration due to their significant measurement errors, cc - GC satisfied “collision" conditions, bh - GC  interacted with central Black Hole.} \\
\end{tabular}
}
\label{tab:GC3}
\end{table}

\newpage

\end{document}